\documentclass[11pt]{article}
\usepackage{amssymb,amsmath}
\usepackage{float}
\usepackage{color,fullpage}
\usepackage{amsthm}
\usepackage{graphicx}   
\usepackage{algorithm}
\usepackage{algorithmic} 
\usepackage{graphics} 
\usepackage{caption}
\usepackage{subcaption}
\usepackage{multirow}
\usepackage{url}
\usepackage{mathdots}
\usepackage{arydshln}

\newtheorem{theorem}{Theorem}

\numberwithin{theorem}{section}
\numberwithin{equation}{section}
\numberwithin{definition}{section}

\begin{document}
\bibliographystyle{unsrt}
\title{Multiple Scattering Media Imaging via End-to-End Neural Network}

\author{Ziyang Yuan\thanks{
Department of Mathematics, National University of Defense Technology,
Changsha, Hunan, 410073, P.R.China. Corresponding author. Email: \texttt{yuanziyang11@nudt.edu.cn}}
\and Hongxia Wang{\thanks{
		Department of Mathematics, National University of Defense Technology,
		Changsha, Hunan, 410073, P.R.China. Email: \texttt{wanghongxia@nudt.edu.cn}}
}
}



\date{}

\maketitle

\begin{abstract}
	\indent
	Recovering the image of an object from its phaseless speckle pattern is difficult. Let alone the transmission matrix is unknown in multiple scattering media imaging. Double phase retrieval is a recently proposed efficient method which recovers the unknown object from its phaseless measurements by two steps with phase retrieval.\\ 
	\indent
	In this paper, we combine the two steps in double phase retrieval and construct an end-to-end neural network called TCNN(Transforming Convolutional Neural Network)  which directly learns the relationship between the phaseless measurements and the object. TCNN contains a special layer called transform layer which aims to be a bridge between different transform domains. Tested by the empirical data provided in\cite{Metzler2017Coherent}, images can be recovered by TCNN with comparable quality compared with state-of-the-art methods. Not only the transmission matrix needn't to be calculated but also the time to recover the object can be hugely reduced once the parameters of TCNN are stable.\\
	$\mathbf{Keywords}$:~~neural network,~~phase retrieval,~~multiple scattering media,~~end-to-end
	
\end{abstract}

\section{Introduction}
~\\
\indent
The light incident on the multiple scattering media will suffer from multiple reflection. Thus when observing the object through the multiple scattering media with coherent light, the speckle pattern displayed on the far side of the scatter isn't similar with the object which can be seen from Figure 1. Because the wavefront interferences with itself destructively when passing through the multiple scattering media, many details about the object get lost. At the same time, the transmission matrix about this complex media is hard to be analyzed and constructed. Moreover, the detector such as CMOS or CCD can only record the intensity of the speckle pattern. As is well known, recovering the object from its phaselesss measurements called phase retrieval is an ill-posed inverse problem. Overall, it's hard to recover image of the object from multiple scattering media. Considering the importance of this problem in applications,  there are a batch of techniques to deal with it such as TOF(Time of Flight) method\cite{Velten2012Recovering}, multi-slice light-propagation method\cite{Waller20153D}, strong memory effect method\cite{Freund1988Memory}, holographic interferometry method\cite{Popoff2010Measuring}, temporally modulated phase method\cite{Cui2011Parallel} and double phase retrieval method\cite{Dr2015Reference}\cite{Rajaei2016Intensity}. Interested readers can refer to \cite{Metzler2017Coherent} for a review.\\
\begin{figure}	
	\centering
	\includegraphics[width=5in]{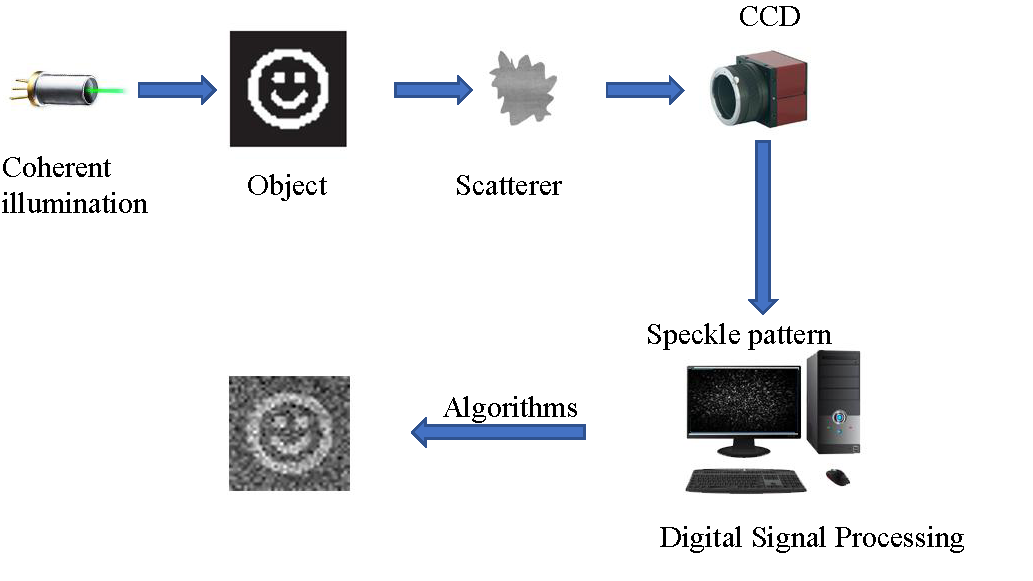}
	\caption{The paradigm of the procedure.}\label{fig:1}
\end{figure}
\indent
Comparing to other methods, the double phase retrieval method can alleviate from the depth and complexities of the scatterer besides this method is able to reconstruct image after capturing only a single speckle pattern once the transmission matrix is estimated. The mathematical formulation of double phase retrieval is:
\begin{eqnarray}
&\mathrm{Find}~\mathbf{A},~\mathbf{x}&\nonumber\\
&s.t.~~|\mathbf{A}^* \mathbf{x}|^2 = \mathbf{b},&
\end{eqnarray}
where $\mathbf{A}\in\mathbb{C}^{n\times m}$ is the transmission matrix, $\mathbf{x}\in\mathbb{C}^n$ is the signal of interest. $\mathbf{b}$ is the measurement, $*$ is the conjugate and transpose. $|\cdot|$ is the element-wise absolute. 
The double phase retrieval method has two main steps: estimating $\mathbf{A}$ firstly, then recovering $\mathbf{x}$ based on $\mathbf{A}$ from the first step. As its name suggested, the core of double phase retrieval is to solve several phase retrieval problems. \\
\indent
Phase retrieval, namely recovering the object from the phaseless measurements which arises from lots of applications such as holographic imaging, coherent diffraction imaging and astronomy,etc. The mathematical formulation of phase retrieval is same with formula (1.1), but the transmission matrix $\mathbf{A}$ is known in advance. The uniqueness of the problem is often up to a global phase factor. During several decades, lots of theoretical analyses and numerical algorithms have come up to solve the phase retrieval problem. Gerchberg and Saxton method \cite{Gerchberg1971A}and Fienup method \cite{Fienup1982Phase} are the two types of classical alternating methods to find the solutions of phase retrieval problem. They have been utilized in various of applications. In \cite{candes2015phase}, it came up with a gradient descent based method called Wirtinger flow to search for the solutions of the non-convex problem.  Theoretical analyses also guarantee the convergence to the global optimum when each column of $\mathbf{A}$ satisfies $\mathbf{a}_i\overset{i.i.d}{\sim}\mathcal{N}(\mathbf{0},\mathbf{I}),i=1,\cdots,m$. Based on this, several works were came up to decrease the sampling complexity besides increasing the recovery probability. Convex relaxation can also be utilized to relief the phase retrieval problem. Phaselift and Phasemax are the two different representatives of convex methods for dimension lifting and constraints relaxation\cite{Cand2013PhaseLift}\cite{Goldstein2018PhaseMax}\cite{Bahmani2016Phasemax}. In theoretical analyses, when $m\geq 2n-1$ or $m\geq 4n-4$ , uniqueness can be guaranteed for phase retrieval when $\mathbf{A}$ is real or generic complex\cite{Balan2006On}\cite{Bandeira2013Saving}. At the same time, oversampling can relief the illness of phase retrieval problem and improve the performance of corresponding algorithms by numerous numerical tests.  

The double phase retrieval can also be called blind phase retrieval. In this case, the transmission matrix $\mathbf{A}$ and $\mathbf{x}$ is unknown. The condition of it is worse than phase retrieval. As a result, in the first step, $\mathbf{A}$ will be evaluated from $k$ known images $\mathbf{x}^{(i)},i=1,\cdots,k$ and their corresponding known intensity measurements $\mathbf{b}^{(i)},i=1,\cdots,k$. In this circumstance, the model can be defined as below:
\begin{eqnarray}
&\mathrm{Find}~\mathbf{A}&\nonumber\\
&s.t.~|\mathbf{A}^*\mathbf{x}^{(i)}|^2=\mathbf{b}^{(i)},i=1,\cdots,k.&
\end{eqnarray}
As a result, define $\mathbf{X}=[\mathbf{x}^{(1)},\mathbf{x}^{(2)},\cdots,\mathbf{x}^{(k)}]$, $\mathbf{B}=[\mathbf{b}^{(1)},\mathbf{b}^{(2)},\cdots,\mathbf{b}^{(k)}]$, each column of $\mathbf{A}$ namely $\mathbf{a}_j$ can be estimated accurately through a series of phase retrieval problems.
\begin{eqnarray}
&\mathrm{Find}~\mathbf{a}_j&\nonumber\\
&s.t.~|\mathbf{X}^*\mathbf{a}_j|^2 = \mathbf{B}^*_j,j=1,\cdots,m,&\nonumber
\end{eqnarray}
where $\mathbf{B}^*_j$ is the $j$th column of $\mathbf{B}^*$. After obtaining the evaluation of $\mathbf{A}$, utilizing standard phase retrieval methods mentioned above, the unknown $\mathbf{x}$ can be recovered from its measurements $\mathbf{b}$. In \cite{Metzler2017Coherent}, they built up different experimental setups to apply multiple-scatter media imaging, then utilizing the double phase retrieval method based on different phase retrieval algorithms to recover the image. Numerous tests show the good performance and robustness of this method.\\
\indent
Though the double phase retrieval method is efficient, it costs lots of computational resources to solve series of phase retrieval problem. It must evaluate the transmission matrix $\mathbf{A}$ firstly which demands plenty of training data and the large computational burdens for solving phase retrieval problem. In the second step, solving the classical phase retrieval method also needs too much computational cost. Besides it seems a little complicated to solve the problem in two steps. Thus time saved, explicit and efficient method need to be devised.

Deep learning has reached much attention since Alexnet won the champion in the ISVRC 2012. Since then, lots of layer structures and optimization methods camp up to accelerate the development of deep neural network. With the aid of the hardwares such as GPU and series of famous open access projections such as Tensorflow, Pytorch and Caffe,  deep learning have been successfully applied in Object detection, autonomous vehicles, Signal processing such as Voice detection and Voice synthetic, inverse problem such as MRI\cite{Wang2016Accelerating}, holography\cite{Jo2017Holographic} and super-resolution\cite{Dong2016Image}. For phase retrieval, there are some previous works which also utilize the neural network to increase the imaging quality. In \cite{Rivenson2017Phase}, they built a neural network to diminish the effect of the twin image. Compared to the classical methods, this neural network only requires the measurements obtained in a single distance besides having a competitive performance. In \cite{Sinha2017Lensless}, it also came up with a neural work to increase the resolution of the image in lensless coherent diffraction image. This neural network is also an end-to-end neural network which can directly transform the diffraction pattern into the image. 

Compared to those works above, the transmission media in our test is worse besides the speckle pattern bear no resemblance to the image. Those factors make it more difficult to recover the object for neural network. In this paper, we combine two steps in double phase retrieval together by deep neural network called TCNN(Transformation Convolution Neural Network) to directly learn the relationship between the intensity of the speckle pattern and the image of object. Applying TCNN into multiple-scatter media imaging, what we need do is to train the neural network which fully utilizes the training sets of double phase retrieval method in the first step. Once the training of network is finished, the recovered image can be obtained immediately by inputting the intensity of speckle pattern into neural network without calculating the transmission matrix. Tests using the multiple-scattering-media imaging data in \cite{Metzler2017Coherent} demonstrate that TCNN can have a competitive performance with the state-of-the-art. When the training procedure of TCNN has been completed, the time required by TCNN is much less than state-of-art for solving the phase retrieval problem when $\mathbf{A}$ is estimated. Besides, TCNN can be refined based on the well-trained network if more training data is available. But double phase retrieval method has to be calculated from all the training data again so that the transmission matrix can be updated. TCNN is a special network which is devised to solve the inverse problem. Thus special structure called transforming layer is constructed in TCNN  which can help the neural network learn the relationship between the transforming domain and object domain so that TCNN can recover the image efficiently .\\
\indent
The reminders of this paper are organized as below. In section 2, the details of the experiment setup and TCNN are given. In section 3, the results of the experiment for TCNN are given. Section 4 is the conclusion. 
\section{The feasible of neural network for solving double phase retrieval}
\indent
Before we describe TCNN, the feasible analyses of neural network to solve double phase retrieval is built. Neural network wants to approximate function $g$ so that $\mathbf{x}$ can be estimated directly from $\mathbf{b}$.
\begin{eqnarray}
g_{\Theta}:\mathbf{b}\rightarrow\mathbf{x},
\end{eqnarray}
 where $\Theta$ is the set of parameters in the neural network which are learned via enough known $\mathbf{b}$ and $\mathbf{x}$. Initially, the existence of the operator $g$ must be discussed.\\
 \indent
 The operator $f:\mathbf{x}\rightarrow\mathbf{b}$ must be  injective or $g$ doesn't exist, where $(f(\mathbf{x}))(i)=|\mathbf{a}_i^*\mathbf{x}|^2$. We can clearly see that when $\mathbf{x}\in\mathbb{R}^n$, $f(\mathbf{x}) = f(-\mathbf{x})$ besides if $\mathbf{x}\in\mathbb{C}^n$, $f(\mathbf{x}) = f(c\mathbf{x})$, $|c|=1$. As a result, the injective of $f$ can be satisfied only if $\mathbf{x}$ defined up to a global phase factor. Thus we consider the map $f:\mathbb{R}^n/\{\pm1\}\rightarrow\mathbb{R}^m$ when $\mathbf{x}$ is real, $f:\mathbb{C}^n/\mathbb{T}\rightarrow\mathbb{R}^m$ when $\mathbf{x}$ is complex(where $\mathbb{T}$ is the complex unit circle). Next we introduce the theorem below to guarantee the injective of $f$. 
\begin{theorem}\cite{Balan2006On}
	Assuming there is no noise, $\mathbf{a}_j, j=1,\cdots,m$ are generic frame vectors, when $\mathbf{a}_j\in\mathbb{R}^n$, $\mathbf{x}\in\mathbb{R}^n$ or $\mathbf{a}_j\in\mathbb{C}^n$, $\mathbf{x}\in\mathbb{C}^n$, if $m\geq2n-1$ or $m\geq4n-2$, $f$ is injective.
\end{theorem}
Because the light passing through the multiple-scattering media which may be frost glass or painted wall. In classical analyses, the rows of transmission matrix $\mathbf{A}$ of those media are often assumed satisfying the gaussian or sub-gaussian distribution which are generic vectors. So the property of transmission matrix in Theorem 2.1 can be satisfied. Before training the network, it must sift the dataset so that every $\mathbf{b}^{(i)},~i=1,\cdots,k$ is different to guarantee injective. 

\indent
Now that the existence of the inverse function $g$ is guaranteed. As a result, we can build up the universality theorem below:
\begin{theorem}
	If the inverse function $g$ of phase retrieval $f$($f:\mathbb{R}^n/\{\pm1\}\rightarrow\mathbb{R}^m$ or  $f:\mathbb{C}^n/\mathbb{T}\rightarrow\mathbb{R}^m$) exists, then $g$ can be approximatd by $g_{\Theta}$ with any desired degree of accuracy.
\end{theorem}
Proof. When $\mathbf{x}\in\mathbb{R}^n$, $f:\mathbf{x}\rightarrow\mathbf{b}$ is a continuous and borel measurable function. As a result, if $g$ exists, $g$ is also a borel measurable function. Utilizing the universality theorem in \cite{Hornik1989Multilayer}, $g$ can be approximated by $g_{\Theta}$ with squashing functions to any desired degree of accuracy as if hidden units are sufficient.\\
\indent
When $\mathbf{x}\in\mathbb{C}^n$, we split $\mathbf{x}$ into two parts $\mathbf{x}_1$ and $\mathbf{x}_2$ uniquely which satisfy $\mathbf{x}_1+j\mathbf{x}_2=\mathbf{x}$ where $j=\sqrt{-1}$. Because $g$ exists, as a result, let $g^{(1)}=(g+\overline{g})/2$, $g^{(2)}=(\overline{g}-g)j/2$, so $g^{(1)}(\mathbf{b})=\mathbf{x}_1$, $g^{(2)}(\mathbf{b})=\mathbf{x}_2$. Because $g^{(1)}$ and $g^{(2)}$ are all borel measurable function. Thus, utilizing the results when $\mathbf{x}\in\mathbb{R}^n$, $g^{(1)}$ and $g^{(2)}$ can be approximated by $g^{(1)}_{\Theta}$ and $g^{(2)}_{\Theta}$. So $\mathbf{x}\approx g^{(1)}_{\Theta}(\mathbf{b})+jg^{(2)}_{\Theta}(\mathbf{b})$.\hfill$\square$
\indent
Above all, utilizing the universality of neural network, when the training set is in $\mathbb{R}^n/\{\pm1\}$ or $\mathbb{C}^n/\{\mathbb{T}\}$, it is possible to train a neural network to approximate this inverse function $g$. In this paper, we build a neural network called TCNN to train $g_{\Theta}$ from dataset.
The core of TCNN is to utilize two sections $g_{\Theta_1}$ and $g_{\Theta_2}$ to approximate the inverse function $g$ which can simulate the imaging procedure in multiple scattering media, namely,
\begin{eqnarray}
g(\mathbf{\mathbf{b}})\approx g_{\Theta}(\mathbf{b})= g_{\Theta_2}(g_{\Theta_1}(\mathbf{b})).
\end{eqnarray}
For $g_{\Theta_1}$, features of intensity of speckle patterns $\mathbf{b}$ are learned from different scales which can be seen from Figure 2. We can see that the details from different scales of speckle pattern are learned(the features in Figure 2 are resized to keep the same). Then deconvolution procedures decode those features into a new element $g_{\Theta_1}(\mathbf{b})$ in the transform domain so that it can be approximated to $\mathbf{x}$ by linear transform $g_{\Theta_2}$. 
\begin{figure}
	\centering
	\includegraphics[width=5.5in]{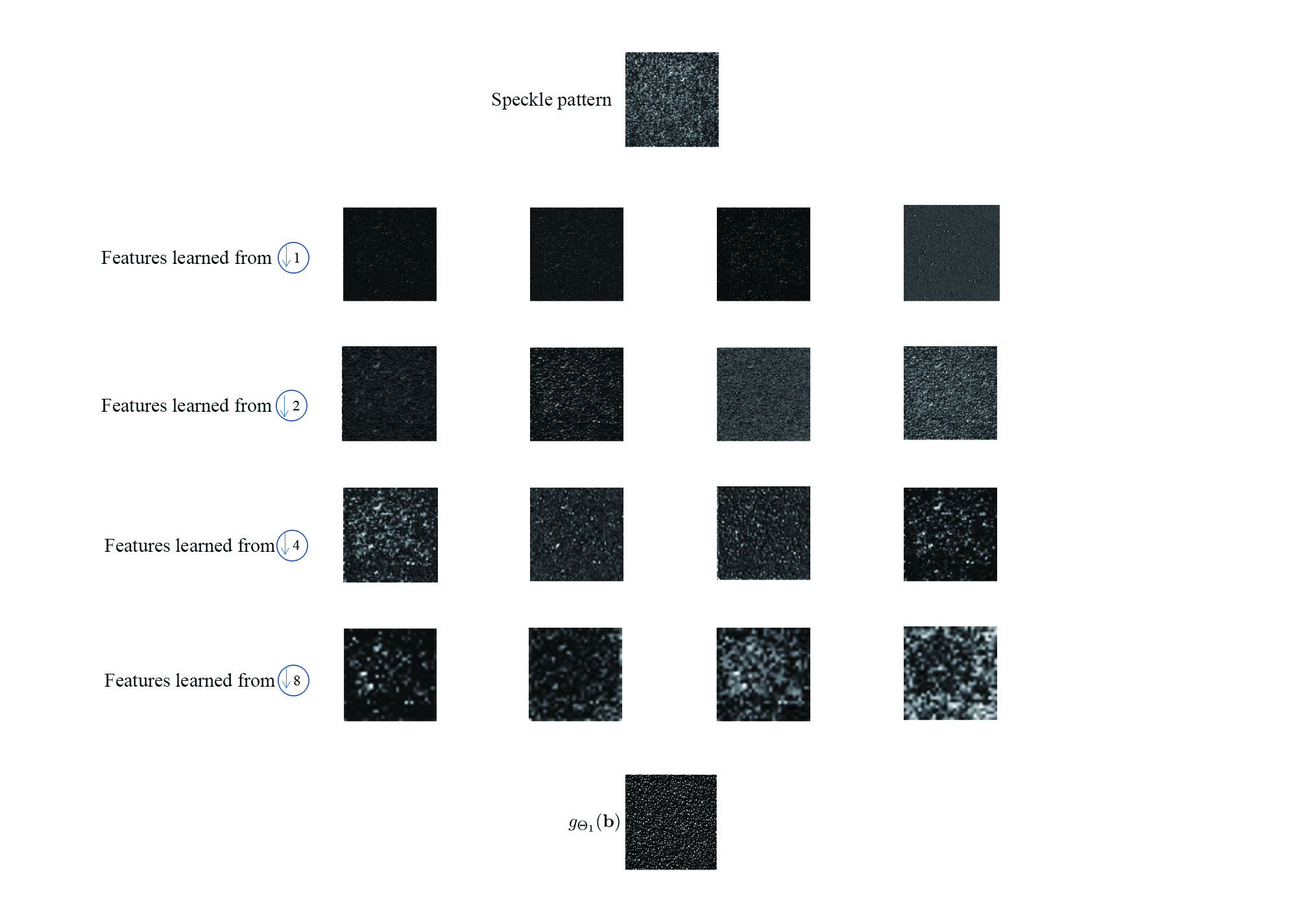}
	\caption{The features learned in TCNN.}
\end{figure}
The diagram of TCNN can be viewed in Appendix.\\
\indent
The multiple flow structure is utilized in TCNN which can learn the features of the input in different scales. The input of neural network is decimated by the \textit{downsample Block} which is constructed by one convolutional layer with batch normalization, Relu activation function and one maxpooling layer. The maxpooling layer here decreases the dimension of the input. Specifically the input is downsampled by $\times2$,$\times4$,$\times8$ respectively. Then, the four different tensors will be successively passed through five \textit{residue blocks}, each block contains two convolutional layers with batch normalization and a shortcut between the input and output of the block. The shortcut can accelerate the convergence of TCNN. After the five \textit{residue blocks}, the high order features in different scales are learned. Then, we will decode those features to generate four different tensors keeping the same size with the input. So there will be $3,~2,~1$ \textit{upsampling blocks} for each tensors respectively. In each \textit{upsample block}, there will be one convolutional layer with batch normalization and one deconvolutional layer. The deconvolution in this paper adopts the way of upsampling in \cite{Rivenson2017Phase} for super-resolution. Because this method can alleviate the effect of zero padding by traditional deconvolution besides fully utilizing the information in the network. After those \textit{upsample blocks}, the fused tensor is obtained which will pass through the \textit{transformation block} instituted by one convolutional layer and transformation layer. The transformation layer is in fact a full connection layer which acts as a linear transformation between the transform domain and object domain. In the test, to alleviate the influence of over-fit, the units in transformation layer are randomly neglected. The details of TCNN can be found in Appendix.
\section{The test of empirical data} 
The experiment setup displayed in this paper is made by Rice University in \cite{Metzler2017Coherent}. Here, we only display the experiment setup for phase only SLM. 
As shown in Figure 3, a spatially filtered and collimated laser beam($\lambda = 632.8nm$) illuminates an SLM from Holoeye. It is a reflective type display(LC2012) with $1024\times768$ resolution and $36$ micrometer size square pixels. It can modulates the phase of the beam before the lens $L(f=150mm)$, this lens can focus the beam onto the scattering medium which is a holographic 5 degree diffuser from Thorlabs. Then a microscope objective(Newport, X10, NA:0.25) is used to image the SLM calibration pattern onto the sensor which is the Point Grey Grasshopper 2 with pixel size 6.45 micrometer. Because the phase only SLM is 8 bit, it modulates the wavefront by an element of $\{e^0,e^{2\pi j\frac{1}{256}},\cdots,e^{2\pi j\frac{255}{256}}\}$. In the test, the phase modulation is restricted to $\{0,\pi\}$, which means the value of the $\mathbf{x}$ is $\{-1,1\}$. For the amplitude only SLM, set the source pixel as either completely off, 0 , or completely on ,1. then the value of $\mathbf{x}$ is $\{0,1\}$. For the details of the experimental setup, readers can refer to \cite{Metzler2017Coherent}.\\
\begin{figure}
	\centering
	\includegraphics[width=3in]{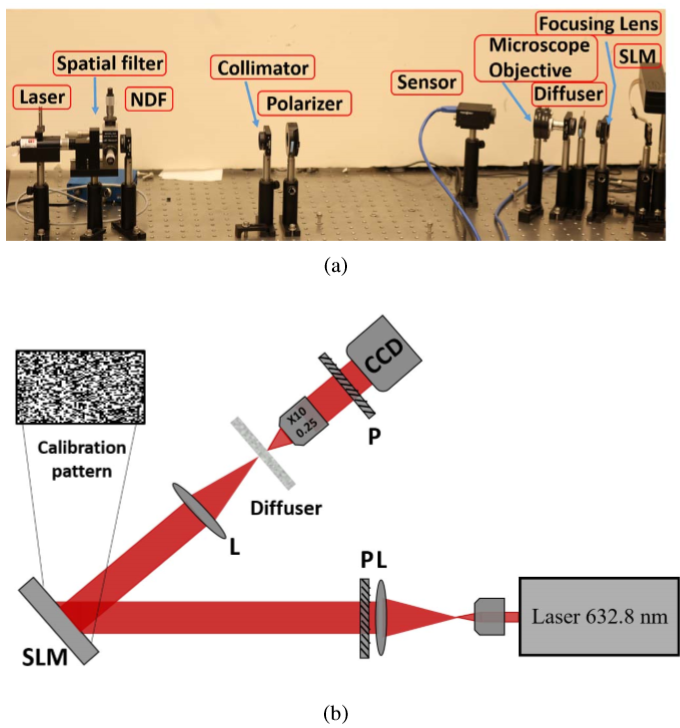}
	\caption{Experimental setup with phase only SLM\cite{Metzler2017Coherent}}
\end{figure}
\indent
The test data can be download from \url{https://rice.app.box.com/v/TransmissionMatrices}. Two types of the data are utilized in the experiment. For amplitude only SLM, the size of the image is $16\times16$ or $64\times64$. For the phase only SLM, the size of the image is $40\times40$. Experiments are applied on desktop computer with GPU NVIDIA 1080 and the CUDA edition is 9.0. Some hyper parameters of the test can be found in Table 1. We imply TCNN using deep learning framework tensorflow. The loss function is chosen as the mean squared error(MSE) between the outputs of TCNN and corresponding benchmarks in the training set. The weights of TCNN are trained by the back propagation using ADAM. The initial learning rate of ADAM is $10^{-3}$ which decays with the factor of $0.85$ after each epoch. The total number of epoch of the training procedure is $100$. To preprocess the data in this experiment, we firstly sift the data so that every image and corresponding speckle pattern is unique. Then we trained TCNN for three different sizes of data. \\
\indent
We compare TCNN with state-of-the-arts in literature such as GS\cite{Gerchberg1971A}, WF\cite{candes2015phase}, PhaseLift\cite{Cand2013PhaseLift}, Prvamp\cite{Metzler2017Coherent}. TCNN can directly recover images from those speckle patterns without estimating the transmission matrix. For other methods, we utilize the transmission matrix $\mathbf{A}$ obtained by Prvamp for phase retrieval. The results are shown in Figure 4 and Figure 5.

\begin{table}[htp]
	\centering
	\caption{Some hyper parameters in TCNN}
	\vspace{0.01cm}
	\scalebox{1}{
		\begin{tabular}{ccccccccc}
			\hline
			\multicolumn{1}{c}{} &\multicolumn{2}{c}{$16\times16$}& & \multicolumn{2}{c}{$40\times40$}& &\multicolumn{2}{c}{$64\times64$}\\
			\hline
			The number of the training sets &\multicolumn{2}{c}{3050}& &\multicolumn{2}{c}{3050}&&\multicolumn{2}{c}{35000}\\
			The number of the validation sets &\multicolumn{2}{c}{22}& &\multicolumn{2}{c}{20}&&\multicolumn{2}{c}{200}\\
			The number of test sets&\multicolumn{2}{c}{5}& &\multicolumn{2}{c}{5}&&\multicolumn{2}{c}{6}\\
			\hline
	\end{tabular}}
\end{table}
\indent
From Figure 4 and Figure 5 we can observe that pictures recovered by TCNN are competitive with state-of-arts. It can recover the outline of the object although some details are lost just like other methods. In fact, it is very hard to recover those images with high quality from speckle patterns. Because multiple scattering media deteriorates the imaging procedure besides the phase information lost by CCD is important to recover image. Moreover, the noise and system error also exist in the experiment. But Table 1 shows the time cost by TCNN is far less than other methods. It can save the time at least ten folds even 100 folds when $64\times64$ which demonstrates the advantages of TCNN to realize real time imaging. Moreover TCNN only performs an end-to-end recovery but other methods must estimate the transformation matrix in advance then getting the evaluation by phase retrieval. Thus, experiments above fully illustrates the power of TCNN to recover the image from the intensity of speckle pattern.\\
\begin{figure}
	\centering
	\includegraphics[width=6in]{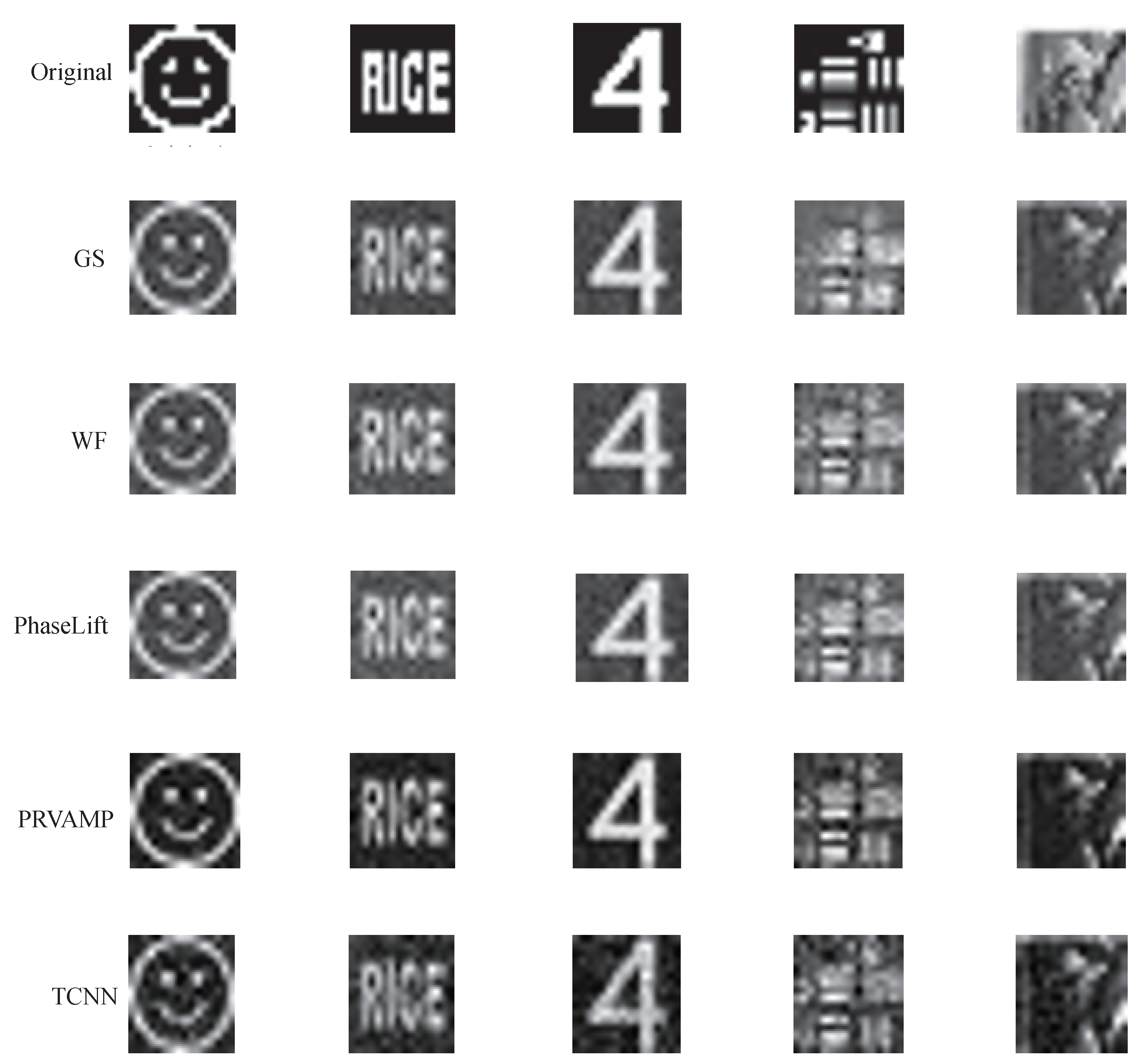}
	\caption{The reconstruction of various methods for amplitude only SLM with image size $16\times16$}
\end{figure}
\begin{figure}
	\centering
	\includegraphics[width=5in]{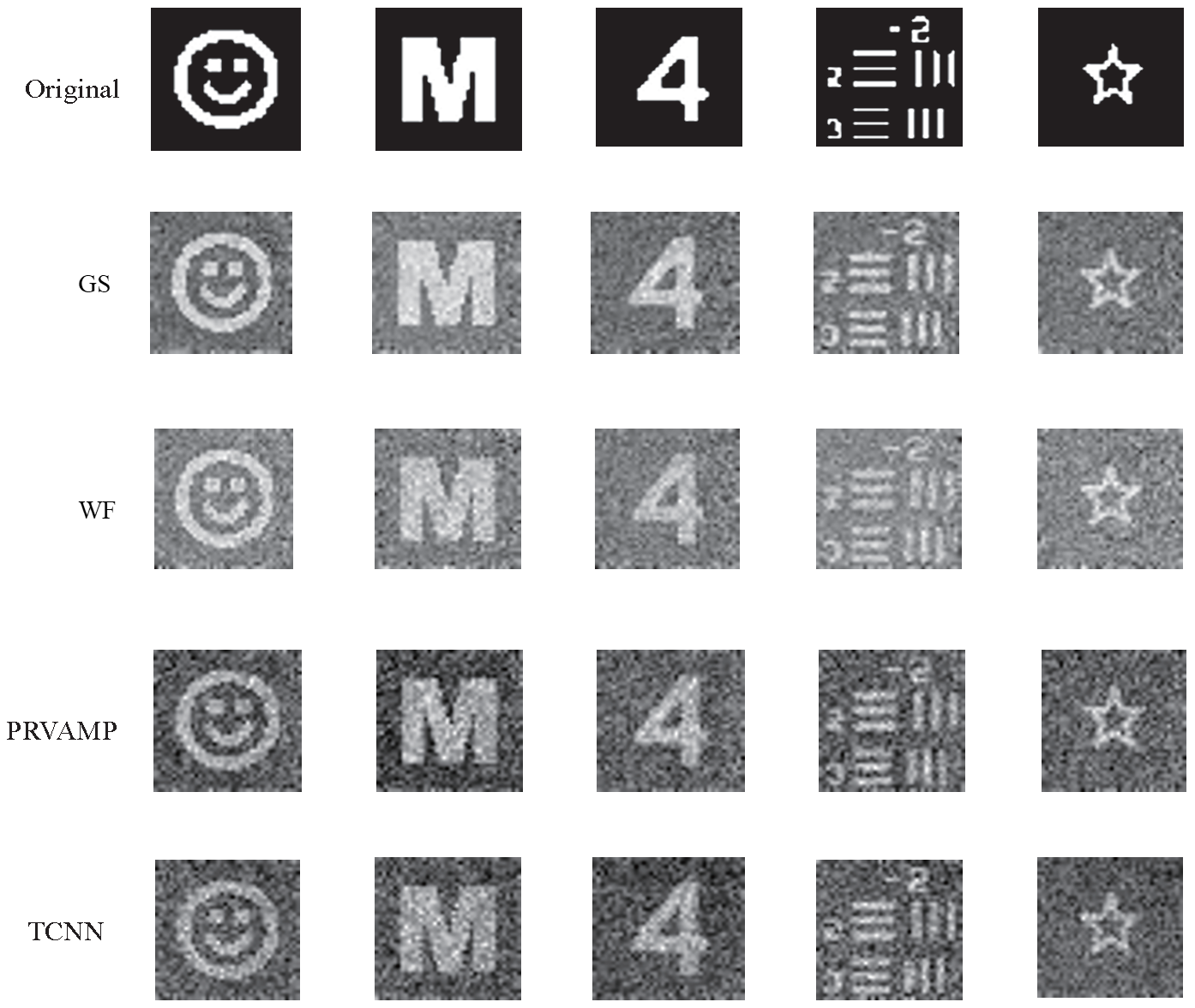}
	\caption{The reconstruction of various methods for phase only SLM with image size $40\times40$}
\end{figure}
\indent
During the training step, the training error and validation error of amplitude only SLM $16\times16$ are given in Figure 6. Considering the large quantities of training set, training error is the MSE of one of the input in training set by random so there is some oscillations for training error. The validation error is the mean MSE for the validation set. From Figure 6, we can find the solution quickly converges to the local optimum, after 20 epochs, there is little fluctuation for both errors. Besides the validation error is very close to the training error which fully demonstrates the influence of over-fit got controlled.
\begin{figure}
	\centering
	\includegraphics[width=3in]{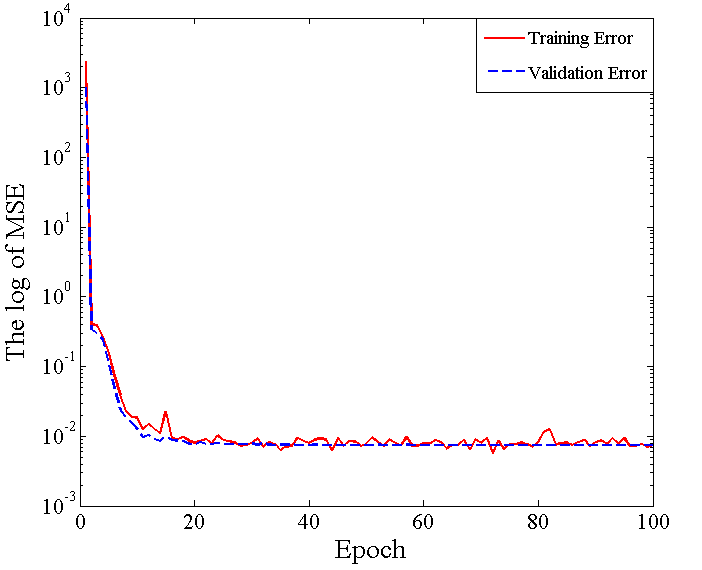}
	\caption{The training error and validation error}
\end{figure}
\begin{table}[htp]
	\centering
	\caption{Time cost per image by different methods}
	\vspace{0.00001cm}
	\begin{tabular}{ccccccccc}
		\hline
		\multicolumn{1}{c}{} &\multicolumn{2}{c}{$16\times16$}&&\multicolumn{2}{c}{$40\times40$}&&\multicolumn{2}{c}{$64\times64$}\\
		\cline{2-3}\cline{5-6}\cline{8-9}
		\multicolumn{1}{c}{}&Time(s)& iteration&&Time(s)& iteration&&Time(s)& iteration\\
		\hline
		WF&$4.11$&$100$&&$19.49$ &$100$&&$289.03$&$100$\\
		GS&$3.86$&$100$&&$18.86$ &$100$&&$43.48$&$100$\\
		PhaseLift&$239.50$&$100$&&---&$500$&&---&$1000$\\
		PRVAMP&$3.41$&$100$&&$17.40$&$100$&&$44.52$&$100$ \\
		TCNN &$0.103$&---&&$0.117$&---&&$0.121$&--- \\
		\hline
	\end{tabular}
\end{table}
\section{Conclusion}
In this paper, we build a deep neural network called TCNN to directly transform the intensity of the speckle pattern via the multiple scattering media to the object. Compared to the traditional double phase retrieval method, this end-to-end network don't need to model this imaging procedure and calculate the transformation matrix. Instead, it needs plenty of training data which includes the original images and corresponding intensity of speckle patterns to update the parameters of TCNN. This training procedure can be done in advance. Compared to state-of-the-arts, the recover quality of TCNN is competitive but the time TCNN cost is much less specifically recovering per image needs no more than $1$s.\\
\indent
In TCNN, we build two sections to approximate the inverse function. The nonlinear part encodes the information of the intensity of the speckle pattern besides decoding it into a new element in transform domain, then the linear part transforms this element into the output in object domain. Test demonstrates the effectiveness of the architecture of TCNN besides tricks utilized in TCNN to alleviate the over-fit of network.\\
\indent
In the future, the work is to decrease the number of parameters in TCNN. Especially for the parameters in the transformation layer, they occupy large portions of the whole parameters. We consider fuse it implicitly into the convolutional layers where the parameters are comparatively less.
\section{Acknowledgement}
This work was supported in part by National Natural Science foundation(China): 61571008.\\
\bibliography{1.bib}


\section{Appendix}
	\begin{figure}	
		\centering
		\includegraphics[width=6.5in]{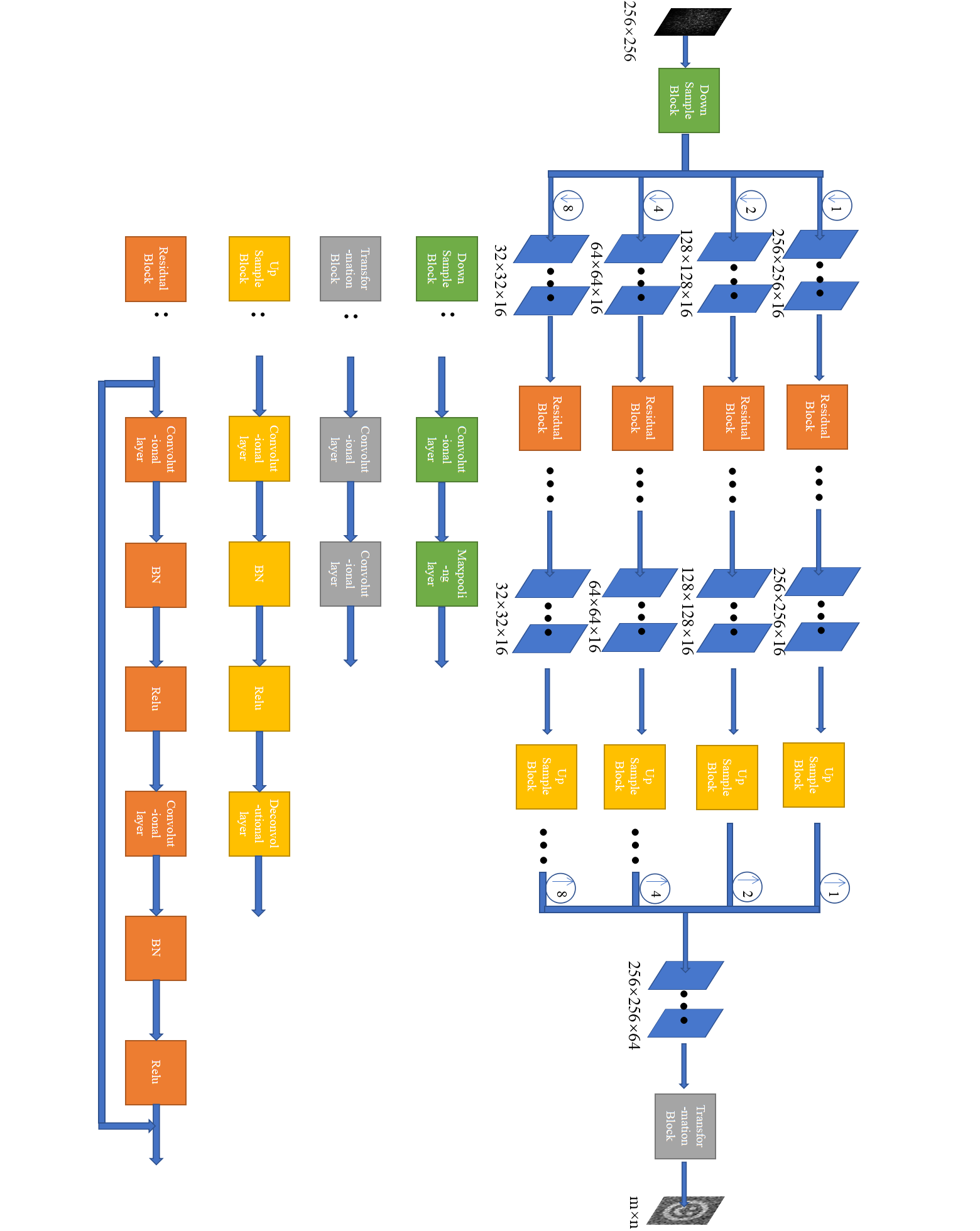}
		\caption{The paradigm of TCNN.}\label{fig:1}
	\end{figure}
\begin{figure}
	\centering
	\includegraphics[width=5in]{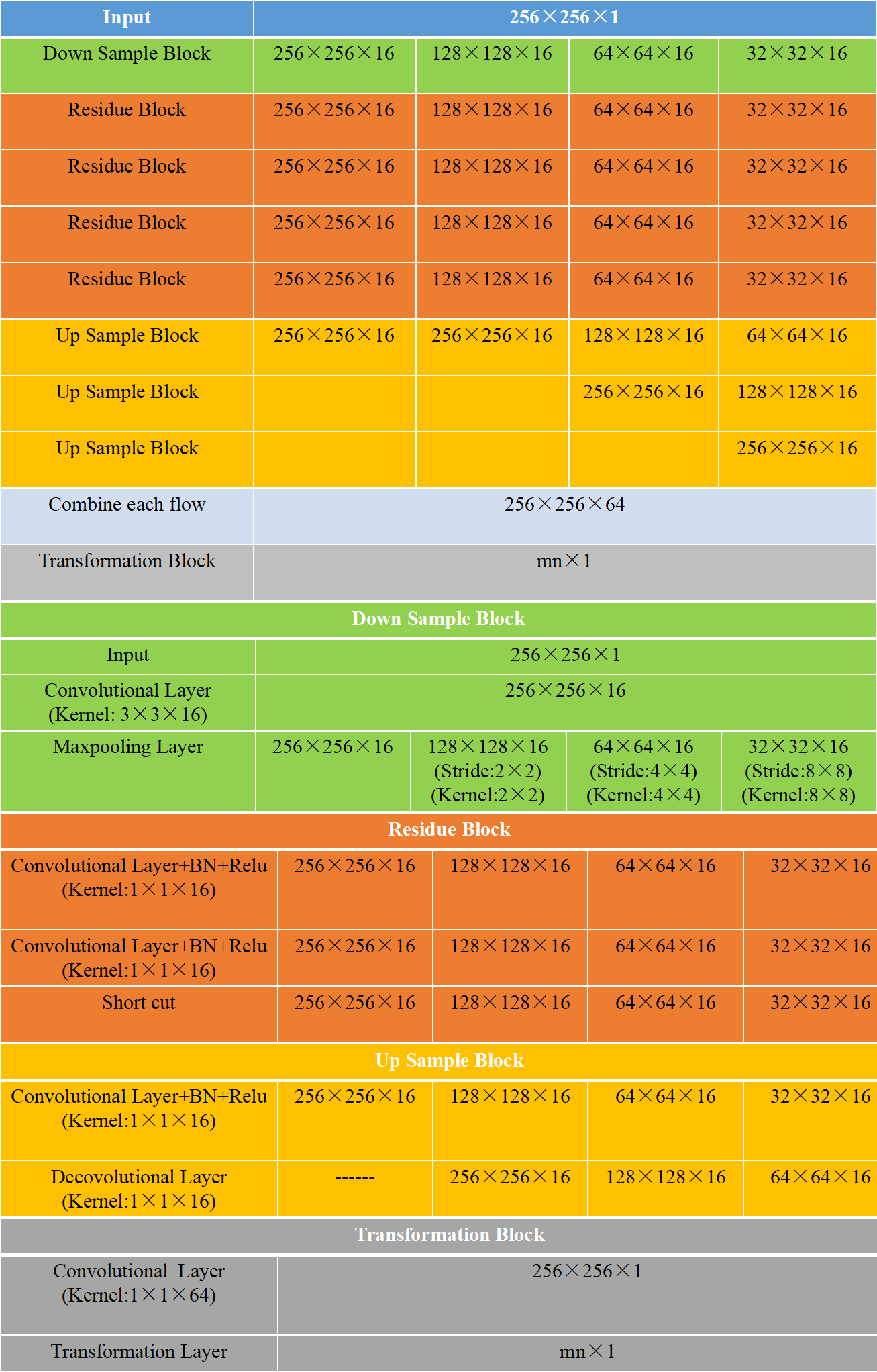}
	\caption{The details of TCNN}
\end{figure}
\end{document}